# An Exploration of Learning Processes as Process Maps in FLOSS Repositories

Patrick Mukala, Antonio Cerone and Franco Turini
Department of Computer Science, University of Pisa
{patrick.mukala, cerone, turini} @di.unipi.it

**Abstract.** Evidence suggests that Free/Libre Open Source Software (FLOSS) environments provide unlimited learning opportunities. Community members engage in a number of activities both during their interaction with their peers and while making use of the tools available in these environments. A number of studies document the existence of learning processes in FLOSS through the analysis of surveys and questionnaires filled by FLOSS project participants. At the same time, the interest in understanding the dynamics of the FLOSS phenomenon, its popularity and success resulted in the development of tools and techniques for extracting and analyzing data from different FLOSS data sources. This new field is called Mining Software Repositories (MSR). In spite of these efforts, there is limited work aiming to provide empirical evidence of learning processes directly from FLOSS repositories.

In this paper, we seek to trigger such an initiative by proposing an approach based on Process Mining to trace learning behaviors from FLOSS participants' trails of activities, as recorded in FLOSS repositories, and visualize them as process maps. Process maps provide a pictorial representation of real behavior as it is recorded in FLOSS data. Our aim is to provide critical evidence that boosts the understanding of learning behavior in FLOSS communities by analyzing the relevant repositories. In order to accomplish this, we propose an effective approach that comprises first the mining of FLOSS repositories in order to generate Event logs, and then the generation of process maps, equipped with relevant statistical data interpreting and indicating the value of process discovery from these repositories.

## 1 Introduction

Over the years, there has been increasing interest in exploiting learning opportunities in FLOSS communities. Numerous studies conducted in this regard suggest the existence of learning opportunities in FLOSS environments [1-9]. It has emerged that collaborative and participatory learning between participants successfully occurs in such environments [5, 7-8]. This has potentially fostered the levels of interest as well as the aura linked with the occurrence of learning within FLOSS, attracting practitioners from tertiary education and urging them to consider incorporating participation in FLOSS projects as a requirement for some Software Engineering courses [3, 7, 10-14]. A number of pilot studies have been conducted in order to evaluate the effectiveness of such an approach in traditional settings of learning [2-5, 14-15].

Furthermore, an emphasis on how learning occurs in terms of phases has been studied [18-19, 29]. To this end, it has been proposed that a typical learning process in FLOSS occurs in three main phases: Initiation, Progression and Maturation. In each phase, a number of activities are executed between Novices and Experts. A Novice is considered as any participant in quest of knowledge while the knowledge provider is referred to as the Expert. A number of activities are performed during the course of interactions between these participants across the three learning phases. These activities can range from formulating a question, observing a discussion, answering a question to developing code, committing changes and reporting bugs.

FLOSS repositories such as CVS, Bug reports, mailing archives, Internet relay chats all contain traces of participants' activities. Sowe and Stamelos [1], Cerone and Sowe [4] as well as Cerone [18] argue that FLOSS environments typically include discussion forums or mailing lists where users can post questions and get help from developers or other users. Forums are unrestricted and act as learning environments for both Novices and Experts. While many studies have provided invaluable insights in this direction, their results are mostly based on either surveys or observation reports [10, 14-16, 21-22] or stochastic or probability studies conducted in FLOSS communities [29]. Singh *et al.* [29] proposed a stochastic model that aims at capturing learning dynamics in FLOSS environments. The main realization of this study was the design of a Hidden Markov Model (HMM) used to investigate the extent to which individuals learn from their own experience and from interactions with peers. Moreover, the model aims to determine whether an individual's ability to learn from these activities varies as the individual evolves/learns over time, and to what extent the learning persists over time [29].

In spite of these few mentioned attempts and many more available in the literature, there has been no work, to our knowledge, conducted on how the traces of data in FLOSS repositories can be mined specifically in order to study learning patterns. Therefore, we propose to contribute in this context by studying learning activities from FLOSS repositories using process mining. We succinctly present a number of steps undertaken as part of our approach for mining and exploring these repositories in order to trace and produce a pictorial representation of learning processes as process maps.

Such a representation plays a pivotal role in fostering the understanding of potential learning processes in FLOSS communities.

The rest of the paper is structured as follows. Section 2 presents a short review of learning processes in FLOSS communities. Section 3 briefly introduces process mapping and explains it in relation to learning processes. In Section 4 we discuss the data collection and analysis techniques pertaining to the construction of the relevant Event logs used for process mapping. In Section 5 we describe the results of our analysis based on the process maps. In Section 6 we further discuss our approach and outcomes with respect to related work and highlight their contribution to the understanding of learning processes in FLOSS communities.

## 2 Learning processes in FLOSS communities: An overview

The bulk of reports on FLOSS members' profiling, such as the ones by Glott *et al.* [21-22] and Gosh *et al.* [37-38], and the analysis performed by Krishnamurthy [39] have found that OSS members in these communities hold different roles that define their responsibilities and participation in the community activities. These include testers, debuggers, project managers, co-developers and the core developers that make up the core development team. Among these roles, project initiators and the core development team remain at the heart of any development project in the community. This is made up of a small number of developers while the rest of contributors, referred to as the enhanced team, perform additional tasks such as feature suggestions, testing and query handling [39].

Apart from FLOSS participants who play roles with direct impact on FLOSS project, we can also distinguish between passive and active users of FLOSS products. Passive users are observers whose only active role is the mere use of the products. Active users are members of the community who do not necessarily contribute to the project in terms of coding, but whose support is made through testing and bug reporting [21-22, 37-38].

Figure 1 suggests a progressive skills development process that can be observed in FLOSS communities [31]. As highlighted by Aberdour [30], participants increase their involvement in the project through a process of role meritocracy. This implies that passive users could move from their state of passiveness to active users, bug reporters until they possibly become part of the core team [21, 30]. All these roles represent crucial contributions required for the overall project quality. However, in FLOSS environments, moving to a higher state is regarded as a reward and recognition of members' abilities and contributions [30]. This is illustrated in Figure 1 where, upon attaining a certain level of expertise and contributive role, users move to roles with more responsibilities. Such role migration is also seen as moving to a higher skill level [30-31].

To this end, it has been proposed that a typical learning process in FLOSS occurs in three main phases: Initiation, Progression and Maturation [20, 30]. In every phase, a number of activities are executed between Novices and Experts. A Novice is considered as any participant in quest of knowledge while the knowledge provider is referred to as the Expert.

In Figure 2, we observe a summary of the deliverables expected from contributors' activities as executed in each phase of the development process as well as the main learning activities as shown on the x axis [18]. Moreover, Figure 2 shows how the different activities as carried out by FLOSS contributors are linked to their maturation through learning stages [18]. Each phase in Figure 2 corresponds to a phase in the learning process. The Understanding, Practicing and Developing phases in Figure 2 respectively correspond to the Initiation, Progression and Maturation phases of the learning process.

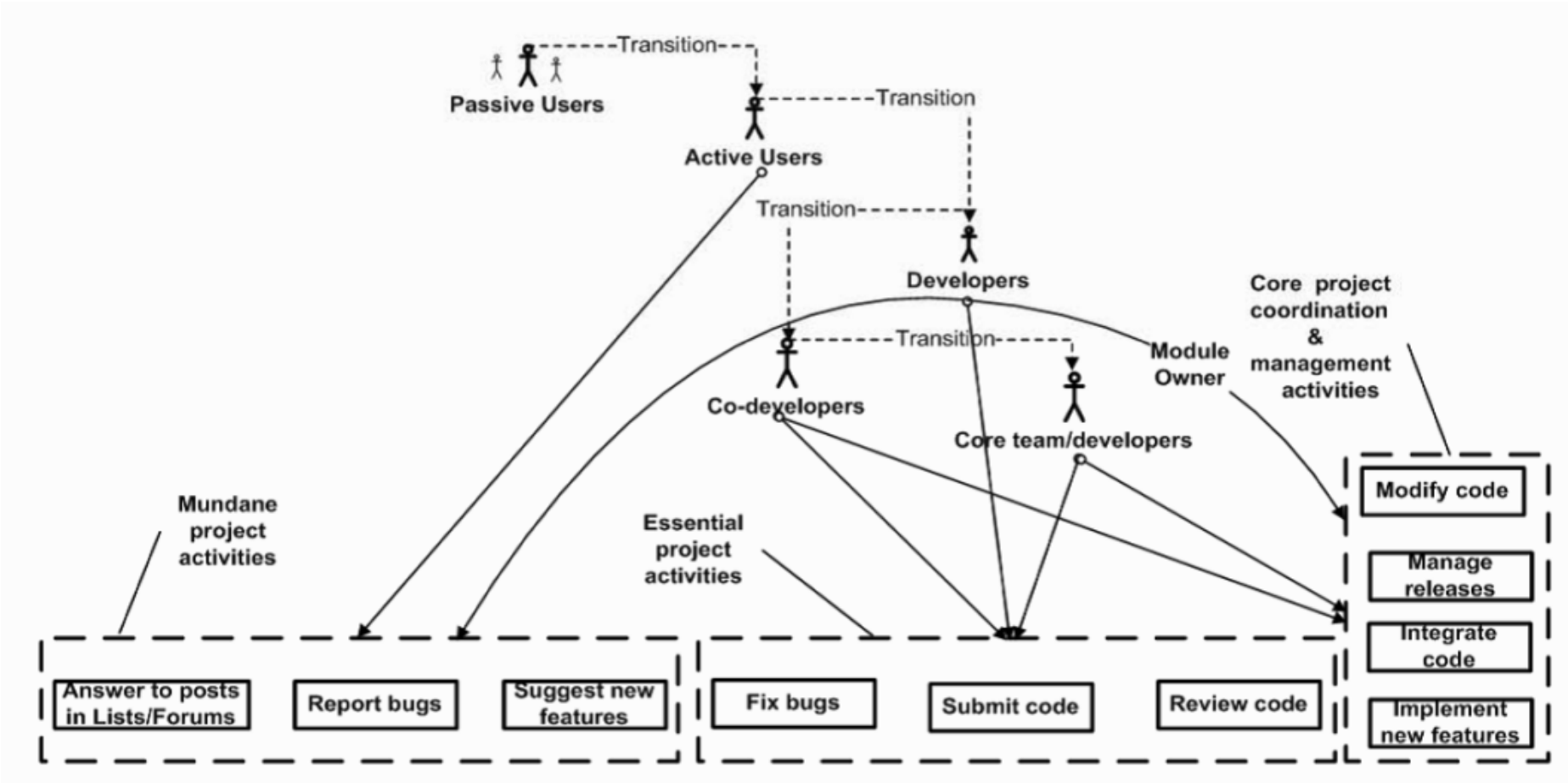

**Figure 1. Role transition in FLOSS communities (From Glott *et al.* [21])**

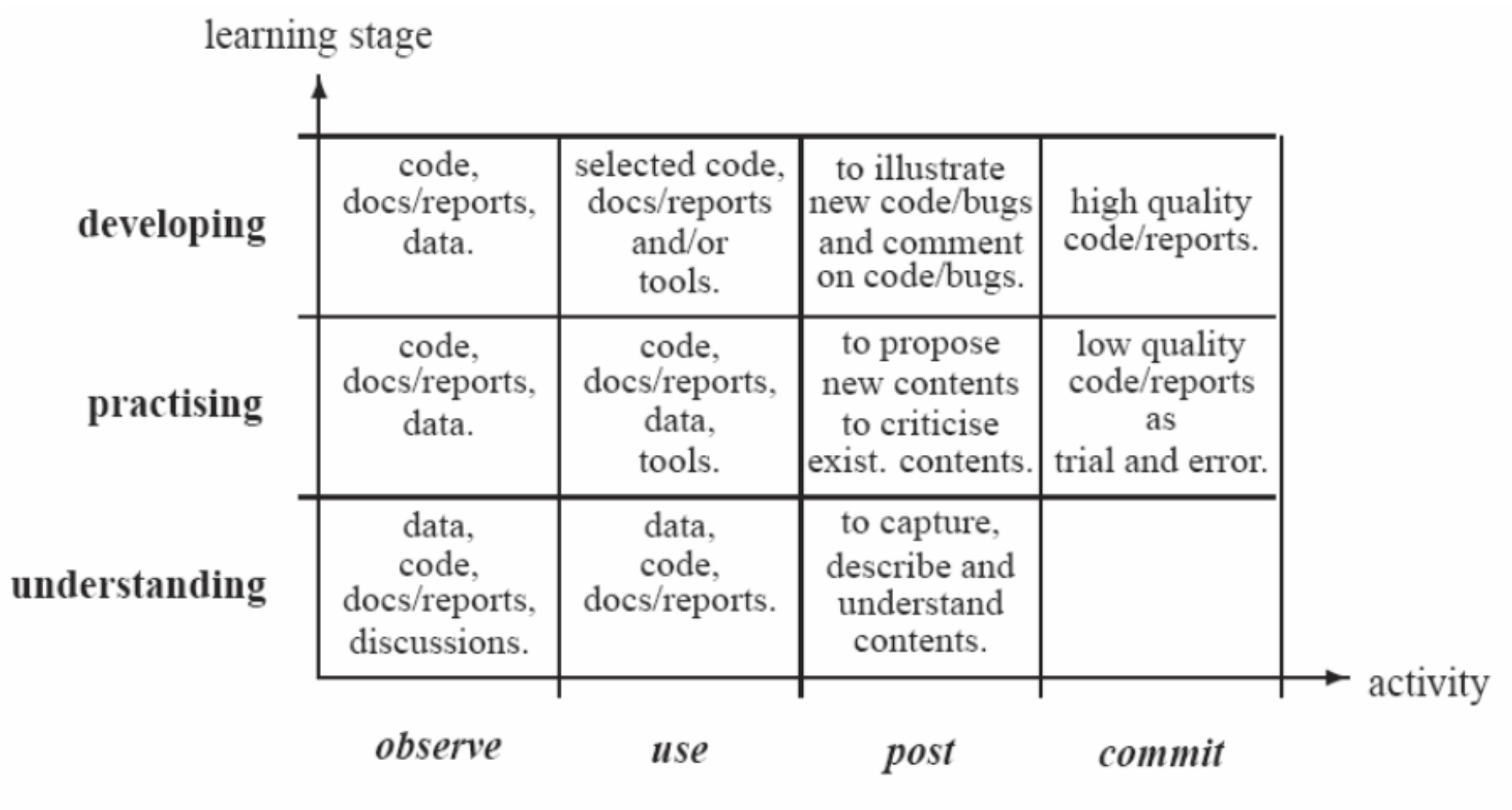

**Figur 2. Learning stages  and Activity contents (From Cerone [18])**

In describing the learning phases we make use of action identifiers (activities) that correspond to events recorded in FLOSS data. The vocabulary and terminology for learning phases in FLOSS are expressed through the OntoLiFLOSS ontology, which was defined in our previous work [42]. The main activities that may be identified through the three phases are described in Sections 2.1 – 2.3.

## 2.1 Initiation

In this phase, both Novice and Expert perform a number of activities that can trigger and perpetuate a learning process.
A Novice seeking help can perform the following activities before making contact with an Expert: *FormulateQuestion* and/or *IdentifyExpert*, and lastly *PostQuestion*, *CommentPost* or *PostMessage*. Then, the Novice can simply perform ContactExpert. If the Expert responds positively, the Novice can perform *SendDetailedRequest* otherwise the cycle for identifying Expert is restarted [20].

The Expert can provide help after performing one of the following activities: *ReadMessages* on mailing lists/Chat messages, *ReadPost* from forums, *ReadSourceCode* as any participant commits code to the project, or *CommentPost*. Just like the Novice seeking to be part of some form of knowledge channel, the Expect can perform ContactNovice and show interest in helping or simply perform *CommentPost*.

## 2.2 Progression

In this phase, both Novice and Expert execute a series of new activities building up from the previous phase.
After accepting a request from the Novice, the Expert performs *ReviewThreadPosts* to be fully aware of the questions and needs for clarification raised by the Novice and *ReviewThreadCode*, for the purpose of critiquing and fixing the code, if needed. The Expert may also perform SendReply in an attempt to answer any direct questions and help requests or just to react to a discussion in a forum. Furthermore, the Expert performs *SendFeedback* and *ReplyPostedQuestion*, to directly or indirectly address doubts or questions from the Novice, *PostQuestions*, to enquiry about possible further needs by the Novice, and *ReportBugs*, as a response to Novice's needs, such as understanding why a piece of code does not run properly.

Moreover, the Expert may monitor the Novice through a set of activities for the purpose of evaluating the level of skill acquisition. These activities include RunSourceCode and *AnalyseSourceCode*, to identify flaws in the Novice's works, and, if necessary, *ReportBugs*, *CommentOnCode* and *ReplyToPost* [20].

The Novice can only react to the Expert's help or feedback by providing insights on the extent to which such help or feedback was useful through *ProvideFeedback*, or simply posing more questions through *PostQuestions*. The Novice also performs a number of activities in the context of posting. These activities may include *PostQuestions*, *ReplyPostedQuestions* and possibly *SendFeedback*.

Furthermore, the Novice can start exercising the new acquired skills through activities such as *AnalyseSourceCode*, when looking at new commits, new pieces of code being posted by community members. Thus, the Novice is able to comment on commits and code through *CommentOnCode* and by reporting bugs through *ReportBugs* [20].

## 2.3 Maturation

In the last phase, Maturation, the learning process intensifies as the knowledge exchanged has matured and allows the Novice to perform more advanced activities according to the progress made in skills accumulation, as depicted in Figure 1.

The Novice is assumed to have acquired enough skills to be able to undertake a set of activities such as *AnalyzeDiscussions* in order to actively engage and contribute to comments and posts about topics in the sphere of the skills acquired and possibly becoming an Expert. Activity *AnalyzeSourceCode* entails looking at the code (when applied) in order to understand and critique that piece of software. Activity *AnalyzeThreadProgression* entails to be part of a discussion and exchange channel that engages on a topic related to a new skill learnt.

Having developed skills in a specific area, the Novice can now commit some deliverables that can be evaluated and criticized by the community. This can be summarized through the following activities: *SubmitBugReport*, consisting in committing any fix or bug report for the interest of the entire community; *SubmitCode*, consisting in committing code; *SubmitDocumentation*, consisting in documentation in terms of requirements elicitation documents, help document, user manuals, tutorials etc. The activity *SubmitCode* is essential in building reputation for a possible role transition.

The Novice also carries out a number of activities that demonstrate core software engineering skills such as developing software, refactoring code, testing and optimizing pieces of code. The activities include: *FixBugs*, consisting in fixing any reported bugs; *GiveSuggestion*, as part of reviewing peers' works and providing alternatives when needed; *PostCommentOnCode*, in order to make sure that appropriate indicative comments that appears in the code are also posted; *ReplyToSuggestion*, to reply and critique suggestions from other Experts or Novices in an active fashion; *WriteSourceCode*, in order to commit pieces of software; *ModifySourceCode*, to modify code and implement suggestions.

The Novice can undertake a number of review activities: *ReviewCommentContents*, in order to contribute to comments and posts about topics in the sphere of the skills acquired and possibly become an Expert; *ReviewPosts* on mailing lists and forums so as to react as needed to comments and posts related to a particular content that is the subject of learning; *ReviewSourceCode*, to explain the ability to analyze the code and identify flaws, which can be reported through activity ReportBugs.

The Expert will perform exactly the same activities while tracking the Novice's progress. These include activities such as *AnalyzeThreadProgression*, *AnalyzeSourceCode* and *AnalyzeDiscussions*. Moreover, the Expert conducts reviews on these activities for monitoring purposes and gives feedback. These activities include *ReviewDocumentation*, *ReviewCode*, *ReviewReport* and *SendFeedback*.

Furthermore, the Expert carries out a number of additional activities as part of the learning process: *RunSourceCode*, *AnalyzeSourceCode*, *CommentOnCode* and *ReportBugs*, and contributes to the Novice's learning process by performing: *ReviewThreadPosts*, to become aware of questions and requests for clarification; *ReviewThreadCode*, for the purpose of critiquing and fixing as required; *SendReply*, as an attempt to answer any direct questions and help requests or just react to a discussion in a forum.

Finally, it is important to note that in this last phase of the learning process, the Expert sometimes performs the same activities as the Novice, but on the Novice's progress: *ReviewPosts*, to react to comments and posts by the Novice, as well as *ReviewSource-Code*, *ReportBugs* and *ProvideFeedback*.

The description herein of the three learning phases provides a broad and generic representation of how communication between FLOSS members also involves learning processes. Such description is critical to understand learning processes in FLOSS environments as it lays the details pertaining to activities and tasks performed by both the Novice and Expert as part of these processes. As pointed out, our aim is to further explore these learning processes by providing empirical evidence directly from data.
The key to such empirical analysis is to make sure that we present a systematic way of mining the FLOSS repositories we have identified and producing the evidence of learning processes in these repositories. Such evidence, in the form of traces accounting for users' activities in FLOSS communities, can be represented through process maps.

## 3 Process Mapping

Process mapping can be considered in simple terms as the step-by-step description of the actions taken by workers (users) as they use a specific set of inputs to produce a defined set of outputs [40]. Literally, a process map is a pictorial representation of the sequence of actions that comprise a process. It provides an opportunity to learn about work that is being performed and a reference to discuss how things get done [32-36].

Process maps can be produced to serve a number of purposes. They can help in outlining processes in various settings: manufacturing processes, corporate structures and management tasks [40]. Process maps provide valuable information about a process to help management find ways to make the process better. The detailed step-by-step descriptions included in process maps provide a clear and concise blueprint for content.

Furthermore, process mapping offers numerous advantages related to managing processes, evaluating and communicating the processes performance and guiding decision making [32-36]. However, for the purpose of this paper, we consider process mapping as a viable means to model and communicate learning processes found in FLOSS repositories. We believe that the visual depiction process mapping offers will reinforce the understanding of learning patters as described through simple surveys or narrative reports. Marrelli [40] argues that process mapping provides a rich and straightforward depiction of processes that requires little or no interpretation. Moreover, process maps include additional elements such as the elapsed time required to perform each step as well as the relative frequency of occurrence for each step (task). This provides critical insights on the scale at which learning occurs in FLOSS communities.

# 4 Data description and analysis requirements

## 4.1 FLOSS repositories for Analysis

The FLOSS platform used for our analysis is OpenStack [23]. According to Wikipedia, "OpenStack is a free and open-source software cloud computing software platform. Users primarily deploy it as an infrastructure as a service (IaaS) solution. The technology consists of a series of interrelated projects that control pools of processing, storage, and networking resources throughout a data center—which users manage through a web-based dashboard, command-line tools, or a RESTful API that is released under the terms of the Apache License" [23].

We considered this platform mainly due to the availability of data about chat messages archives, Source Code and bug reports that we need to track possible learning activities for all three phases of the learning process. Principal activities in the Initiation phase of the learning process are generally about observing and making contacts. Ideally, this phase constitutes an opportunity for the Novice to ask questions and get some help depending on the requests while the Expert intervenes to respond to such requests. In the Progression phase, both participants step up their engagement and commitment during their interactions. The Novice applies the guidelines provided by the Expert who executes a number of activities to ensure such guidelines are implemented. Hence, the Internet Relay Chat (IRC) messages data set is appropriate to trace these activities in the first two phases.

In the Maturation phase, as someone's skillset has matured, we identify a Novice as anybody who is still not acquainted with some topics or parts of the code and requests information about them, while the Expert is respondent to such requests. In order to trace these advanced skills, we look at activities performed by participants who tend towards the last layer of contributors, the core developers as described in Section 2. Such activities can be identified by looking at repositories that store data about developing, creating code, examining and reviewing the code, identifying and fixing possible bugs. Therefore, there are three potential repositories from Openstack that can provide such information: Source Code, Bug reports and Reviews datasets.

However, for the purpose of our analysis, we provide insights considering solely the Source Code dataset. In Source Code, the emphasis is about getting the feeling about how participants react to pieces of code as these are committed, how far they personally contribute in terms of code submission, reviewing code that has been submitted as well as providing feedback with regards to any comment or question asked in the process.

### 4.2 Data description

The Internet Relay Chat (IRC) messages repository contains more than 5 million chat messages, exactly 5603302, exchanged between 19247 people on a combined total of 30 channels/forums. From these chats messages, we eliminated those that could not be linked to senders. The final dataset was made up of 2142690 chat messages that we analyzed. These chat messages were exchanged over a period of 3 and half years. The first message was sent on 2010-07-28 at 05:09:11 and the last message we considered was exchanged on 2014-04-09 at 18:07:19. In this dataset, the average length of a message sent was 60 characters; the longest chat message reached 502 characters while the shortest message was of one character length.

The Source Code repository contains 93584 source code files that are reported to be committed. This is achieved by 2677 people. These people performed a number of actions and these amount to 425744 on about 210 projects. The files submitted are of 75441 types. The different file types were used to identify whether a file committed was documentation, user manual or patch or even any general source file. These files were submitted during a period of time spanning from 2010 to 2014. About 131556 messages can be identified as related to the committed files. The first message recorded was at 23:05:26 on the 27th of May 2010 while the last message included in this analysis was sent at 12:27:48 on the 6th of May 2014. Furthermore, the length of the messages considered is of an average of 182 characters, the longest description/report was of 8628 characters and the shortest message is just one character.

### 4.3 Constructing Event logs

Constructing the Event logs entails retrieving data from all of our repositories as they relate to users' activities, the commit time, the user's details and all elements needed to constitute a full event. In order to be able to deduct that a certain learning activity took place after looking at an annotation or comment posted online, we need to be able to gauge the semantic and contextual meaning of all messages. Therefore, we make use of semantic search with MSSQL as a proper means to achieve this goal.

Semantic search improves search by understanding the contextual meaning of the terms and tries to provide the most accurate answer for a given text document. However, this also requires the use of key phrases to steer the search [24]. Our choice of key phrases is based on a number of studies conducted in FLOSS with regards to the kinds of questions and answers that are asked in FLOSS communication environments [25-29]. We start from this categorization, following questions and responses categories; then we deduct a number of key phrases.

Considering previous findings [25-28], a formal model of learning activities in FLOSS communities [19], as well as lexical semantics, guided by our ontology for learning in FLOSS communities, OntoLiFLOSS [42], we design a set of three catalogs of key phrases that clearly summarize all identified key phrases, the corresponding learning activities as well as the learning phase. Lexical semantics builds from synonyms of terms and their homonyms to derive the meaning of words in specific contexts. The use thereof is paramount and promises to capture the meaning of annotations and messages found in our datasets. It is also critical in identifying the learning activities as we described them in Section 2.

Due to space constraints, we show in Figure 3 only the catalog that contains the key phrases that semantically identify activities as categorized according to the participants' roles in the Initiation phase of the learning process.

In order to produce process maps, we need to identify events that will make up our Event log. An event is a tuple made up essentially of an event identifier, the performer, activity and any other attributes that might be needed. In our case, we include the following attributes: state, date as well as the role (Novice, Expert). An event *E* is a sextuple (*t, a, p, d, s, r*) such that *t* is the case in the event and can be either a topic on emails or an issue number on code and bug reports; *a* is the activity; *p* is the participant; *d* is the relevant date of occurrence; *s* is the state of the learning process phase; and *r* is the participant's role in the process.

Making use of our key phrase catalogs, we then construct the relevant Event logs by retrieving the mappings between key phrases, activities, states and participants. Let $c_1$, $c_2$ and $c_3$ denote respectively Initiation, Progression and Maturation. We distinguish between key phrases for activities and states. Therefore, we refer to key phrases for states as *gl_key* (global keys) while the key phrases that help distinguish activities are referred to as *lc_key* (local keys).
We set catalogs as sextuples ($C, c_i$, *gl_key, state, lc_key, activity, role*) such that *C* is the set of all our catalogs; $c_i \in C$ is a single catalog; *gl_key* is the key phrase for the identification of a state; *state* is the state as it appears in the catalog; *lc_key* is the key phrase used to identify an activity; *activity* is the corresponding activity in the catalog and, finally, *role* is the role as it appears in the catalog. Hence, our final Event logs are constructed using this information as described by the pseudocode shown in Figure 4.
A snapshot depicting an Event log is shown in Figure 5. This Event log has 6 attributes including the name of the contributor, the executed activity, the state of learning phase, the channel topic, the corresponding event date and the role, which in this case is Novice. The channel topic in this log might represents a file number, a question thread or any relevant topic upon which a discussion or debate is based.

| STATES | GLOBAL KEYWORDS | PARTICIPANTS | ACTIVITIES | KEYPHRASE/CONDITIONAL ACTIVITY |
|---|---|---|---|---|
| Observation | "problem", "help", "error" | NOVICE | FormulateQuestion | If PostQuestion = true |
| | | | IdentifyExpert | "How did you do this", "I saw your code", "I need your help", "this does not work for me", "-1", "is this possible to do this", "can this be done?", "very helpful", "very well" |
| | | | PostMessage | If IdentifyExpert = true |
| | | | PostQuestion | "How can I do…?" "How to?", "don't understand how", "could help?", " what is wrong?", "my code is not running", "code not executing", "question", "How to", "what is wrong", "where can I", "Any ideas how to solve this problem?", "I have tried doing", " search for this", "but have had little luck", "any help?", "any suggestions?", "everything I could", "new to the" |
| | | | CommentPost | "does not work", "not executing", "this does not work for me", "do not know what is wrong with my code", "here is my code", "in short my problem", "step by step", "details provided", "works as follows", "I want it to", "expect it to" ,"my question is like this", "what I mean" |
| | | EXPERT | ReadMessages | If CommentPost = true |
| | | | ReadPost | If CommentPost = true |
| | | | ReadSourceCode | "syntax error", "maybe you should…", "it seems to work for me""do not know what is wrong with the code" or "not sure it can"') or "running your code" |
| | | | CommentPost | "system details needed", "more details needed", "more problems details needed", "more details needed of what is on the screen", "Did it work before?", "provide exact step by step details" |
| ContactEstablishement | "can I get your help", "can you help", "send question", "contact details", "send email", "send file", "more details" | NOVICE | ContactExpert | If SendDetailedRequest = true |
| | | | SendDetailedRequest | "actually the code is like this…" "I tried this", "I don't know how"," I don't understand how?" "can you help", "your help", "you explain","as you asked","so my question is", "I wanted to know","what I meant is","my screenshot looks","I get this error","how do I fix this" |
| | | EXPERT | ContactNovice | If CommentPost/SendFeedback = true |
| | | | CommentPost/Send Feedback | "does not work", "not executing", "maybe you should…", "it seems to work for me", "this does not look right" "you need to delete this..", "the syntax is not correct", "send me your code", "what is your problem?", "this works for me", "Did it work before", "I think it should work" |

**Figure 3. Catalog of keyphrases for the Initiation phase of the learning process**

## 4.4 Tool for Process Mapping

In order to process mine these Event logs, we choose an appropriate tool for analyzing the identified events and that can provide efficient visualizations to demonstrate the workflow of occurrence of activities in these processes. We consider Disco [41]. Disco stands for Discover Your Processes. It is a toolkit for Process Mining that enables the user to provide a preprocessed log specifying case, activities, originator and any other attributes. The tool performs automatic process discovery from the log and outputs process maps as well as relevant statistical data

In essence, Disco applies Process Mining techniques in order to construct process models based on available logging data that are turned into an Event log. The logging data contain all the details about transactions that can be found in Event logs. Based on these logs, Disco provides process maps which can be represented with two different set of metrics on the basis of which the flow of events is explained. These metrics include frequency and performance.

The main objective of the frequency metrics is the depiction of how often certain parts of the processes have been executed. We can distinguish three levels of frequency: absolute frequency, case frequency and maximum repetitions. The roles of such levels of frequency can be summarized as follows:

- Absolute frequency: When this value is associated with an activity, it indicates how many times in total that activity has been performed, while it also gives an indication of how often activities moved from one point to another on the edge (path), or how often that particular path has been “travelled” on throughout the whole process;
- Case frequency: This allows to ignore repetitions that might have occurred with activities and only show relative numbers of how many cases passed through which activities and along which path (regardless of whether they came by there just once or multiple times);
- Max. Repetitions: This provides the maximum number of repetitions within a case.

Unlike frequency, the performance metrics provide details about the time at different levels and paths during the execution process. This can also be aggregated to three different levels:

- Total duration: This is the default metric when one needs to display processes according to their performance. It represents the accumulated durations (summed up over all cases) for the execution of each activity and for the delays on each path;
- Mean duration: This is the average time spent within and between activities;
- Max. Duration: This is an indication of the largest execution times and delays that were measured during the process execution.

These metrics can also be provided at once for a path of an activity if needed. It is crucial to note at this point that the process maps we consider in our work represent the activity flows for Novice's and Expert's roles.

```
PROCEDURE: ConstructLog (c_i, c_2, c_3)
Set g_total ← COUNT(gl_key)
Set l_total  ← COUNT(lc_key)
Set a' a derived activity
Set E' a derived event, s' derived state, r' a derived role
Do While chat message, source code, file  still in data set
        For i in 1 to g_total
                If message_body or code annotation  matches gl_key  then
                For j in 1 to l_total
                        If message_body matches lc_key  then
                                t← topic/file ref     ---Get the topic or reference number
                                a← activity           ---Return relevant learning activity
                                p← sender             ---Return the send's details
                                d← date               ---Return the timestamp
                                s← state              ---Return the corresponding state
                                r← role               ---Return the corresponding role
                        Add E (t, a, p, d, s, r) to Event_log   ---add this event to the log
                        Endif
                Endfor
                Else
                                a' = 'Participate in Discussions'  ---A non-identifiable  activity
                                s' = 'Participation'
                                r' = 'Inactive'
                                E' = (t, a', p, d, s', r')
                                Add E (t, a, p, d, s, r) to Event_log
                EndIf
        Endfor

Enddo
END PROCEDURE
```

**Figure 4. Step by step Pseudo Code Rules for Constructing Event logs**

| | Participant | Activity | State | channeltopic | submitted_date | Role |
|---|---|---|---|---|---|---|
| 1 | vish1 | Comment Post (N-O) | Observation | 5 | 2010-07-28 06:33:15.0000000 | Novice |
| 2 | vish1 | Post Message (N-O) | Observation | 5 | 2010-07-28 06:33:15.0000000 | Novice |
| 3 | vish1 | Identify Expert (N-O) | Observation | 5 | 2010-07-28 06:33:15.0000000 | Novice |
| 4 | mtaylor | Contact Expert (N-Es) | ContactEstablishment | 5 | 2010-07-28 06:57:22.0000000 | Novice |
| 5 | mtaylor | Send detailed request (N-Es) | ContactEstablishment | 5 | 2010-07-28 06:57:22.0000000 | Novice |
| 6 | vish1 | Comment Post (N-O) | Observation | 5 | 2010-07-28 09:49:43.0000000 | Novice |
| 7 | vish1 | Post Message (N-O) | Observation | 5 | 2010-07-28 09:49:43.0000000 | Novice |
| 8 | vish1 | Identify Expert (N-O) | Observation | 5 | 2010-07-28 09:49:43.0000000 | Novice |
| 9 | silassewell | Comment Post (N-O) | Observation | 5 | 2010-07-28 14:05:07.0000000 | Novice |
| 10 | jsmith | Contact Expert (N-Es) | ContactEstablishment | 5 | 2010-07-28 14:08:06.0000000 | Novice |
| 11 | jsmith | Send detailed request (N-Es) | ContactEstablishment | 5 | 2010-07-28 14:08:06.0000000 | Novice |
| 12 | dendrobates | Comment Post (N-O) | Observation | 5 | 2010-07-28 14:26:05.0000000 | Novice |
| 13 | dendrobates | Post Message (N-O) | Observation | 5 | 2010-07-28 14:26:05.0000000 | Novice |
| 14 | dendrobates | Identify Expert (N-O) | Observation | 5 | 2010-07-28 14:26:05.0000000 | Novice |
| 15 | silassewell | Contact Expert (N-Es) | ContactEstablishment | 5 | 2010-07-28 14:28:15.0000000 | Novice |
| 16 | silassewell | Send detailed request (N-Es) | ContactEstablishment | 5 | 2010-07-28 14:28:15.0000000 | Novice |
| 17 | creiht | Comment Post (N-O) | Observation | 5 | 2010-07-28 14:57:30.0000000 | Novice |
| 18 | creiht | Post Message (N-O) | Observation | 5 | 2010-07-28 14:57:30.0000000 | Novice |
| 19 | creiht | Identify Expert (N-O) | Observation | 5 | 2010-07-28 14:57:30.0000000 | Novice |

**Figure 5. Snapshot of Event log for Novice during Initiation phase**

# 5 Empirical Results

Taking a FLOSS environment called Openstack as data source, we apply our proposed techniques to identify learning activities based on key phrases catalogs and classification rules expressed through pseudo code as well as the appropriate Process Mining tool. We thus produce Event logs that are based on the semantic content of messages in Openstack Internet Relay Chat (IRC) messages and Source Code to retrieve the corresponding activities. Considering these repositories in line with the three learning process phases (Initiation, Progression and Maturation), we produce an Event log for each category of participant (Novice or Expert) in every phase on the corresponding dataset. Hence, we produce 6 Event logs that help build 6 corresponding process maps which are visual representations of the flow occurrence of learning activities in FLOSS for each category of participant.

We describe these process maps for each category of participant by considering case frequency for the first array of metrics as well as the average duration in order to provide insights within the bounds of our experiments.

## 5.1 Process Maps for the Initiation Phase

After mining the IRC data set, we produce the resultant process maps depicting graphical representations of the occurrence of learning activities in these IRC forums by category of participant.

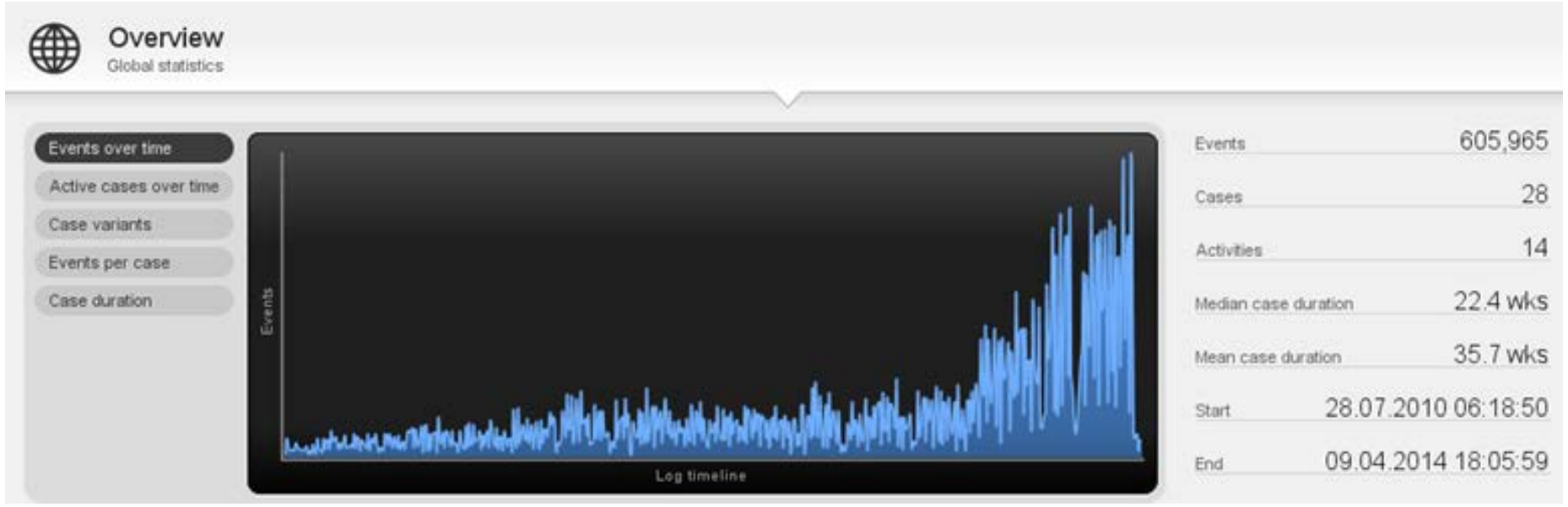

**Figure 6. Event of time on IRC for Initiation Phase**

Figure 6 shows a representation of the time frame for the occurrence of our process events is given. The right hand side of the figure gives some details with regards to input data used in the plot. The log timeline on the horizontal axis represents the total timeframe covered by our dataset (from earliest to latest timestamp observed). We can note that the Initiation Phase of the learning process started on the 28th of July 2010 and ended on the 9th of April 2014. We note that, during this time, a total of 605965 events were generated. An event represents a tuple made up of the case (in this context, the discussion topic), the chat message senders as well as the relevant learning activities. With about 28 cases, a total of 14 activities are executed with an average time per case of 35.7 weeks while the median duration is of 22.4 weeks.

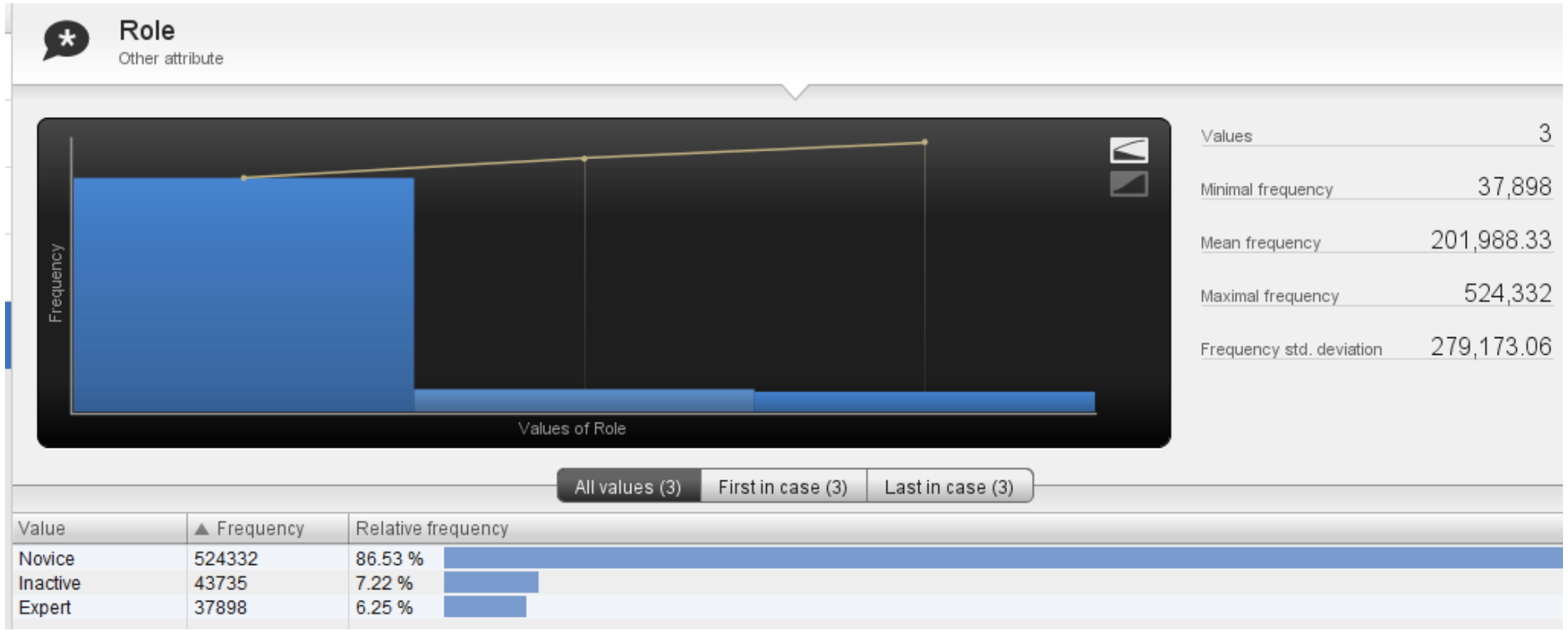

**Figure 7. Details about learning Participants**

Finally, Figure 7 shows how participants in quest for knowledge justifiably claim the majority of activities with a total of 524332 amounting to 86.53% of all executed activities at this point in contrast with Experts who intervene at a lower rate of 6.25% with 37898 activities, slightly behind activities related to doing something other than exchanging knowledge with 43735 activities.

We justify this discrepancy in the next paragraphs as we unpack activities flow for Novices and Experts respectively. As indicated earlier, we shall consider only case frequency for the first array of metrics as well as the average duration in order to provide insights within the bounds of our experiments.

The process map depicted in Figure 8 represents a workflow for all the activities performed by the Novice during the Initiation Phase. The numbers, the thickness of the arcs or edges, and the coloring in the model illustrate how frequently each activity or path has been performed. One can notice that the thickness of the edges is the same. We considered case frequency to identify how many cases were involved in constructing the workflow. In this instance, all 28 cases are considered. The degree to which people engage in sending instant messages and the motivation therein can justify the presence of the workflow in almost every channel.

Cerone [18] hints that an *observe* activity may lead to communication between FLOSS members and that a *post* activity may trigger a learning experience. In Figure 8, the process map depicts the activity flow in terms of learning for a Novice in IRC chats. We note that in 18 cases, the learning process for the Novice starts with commenting a post while in the remaining cases, the Novice triggers this phase through formulating a question. The map demonstrates that the activity *CommentPost* occurs in all 28 cases, although it is the starting point in only 18 of those. From commenting on a post, the Novice would post a message and then identify an Expert. In some cases, the Novice would then formulate an appropriate question to express a request and post the question. In other cases, the Novice would either go back to post further comments on posts or make a direct contact with an Expert by sending a detailed request through formulating a question appropriately. From formulating a question, the Novice can follow 2 paths: either return to *CommentPost* and follow the initial paths or contact the Expert again and continue the indicated paths accordingly.

The existence of such process maps in IRC messages is also highlighted by additional performance indicators. Figure 7 shows that the Novice was active in a total of 524332 events as indicated in the main model performing a total of 7 activities over an average period of 35.7 weeks. The results also indicated that 7 activities were executed on average 74904.57 times with the most executed activity occurring 131888 times, although this is not shown in Figure 7.

Figures 10 and 11 provide insights on the Expert role. The process map in Figure 10 depicts the level of involvement of the Expert during the Initiation Phase. We considered case frequency to identify how many cases were involved in constructing the workflow. The Expert process map in Figure 10 indicates that the Expert gets involved in the learning process by contacting a Novice, commenting a post and probably contacting the Novice again. Then, the Expert reads the source code especially if the Novice's request pertains to coding, and then makes another contact with the Novice probably for further clarification or follow-up comment and follows the same path. Thereafter, the Expert reads messages and posts and then comment on posts before going back to running the code again.

Figure 11 shows that the Expert was active in a total of 37898 events, performing a total of 6 activities over an average period of 35.3 weeks in 28 processes. Our experiment also indicates that these activities were executed on average 6316.33 times with the most executed activity occurring 19506 times.

Therefore, one can observe that learning activities do occur at a significant rate during messages exchange on these forums and the results from mining IRC messages provide insights on the starting activities that trigger these processes. While the dataset boasts over 5 million messages, only 2142690 of these were good enough for our analysis. It has emerged that 14 of the learning activities could be identified through 605965 events across 28 cases where the Novice accounts for the biggest share of traffic during these exchanges. As illustrated in Figure 7, the Novice executes the activities about 86.53% of the total number compared to the Expert who only operates 6.25%.
Before we discuss the second phase, it is crucial to point out that existing literature on learning in FLOSS has not provided insights that would indicate how the communication occurs, the sequential occurrence of learning activities and, more importantly, the level of occurrence of learning activities in FLOSS communities. This level of occurrence could be expressed by looking at the number of learning activities proportionally to the total number of activities that are recorded in repositories.

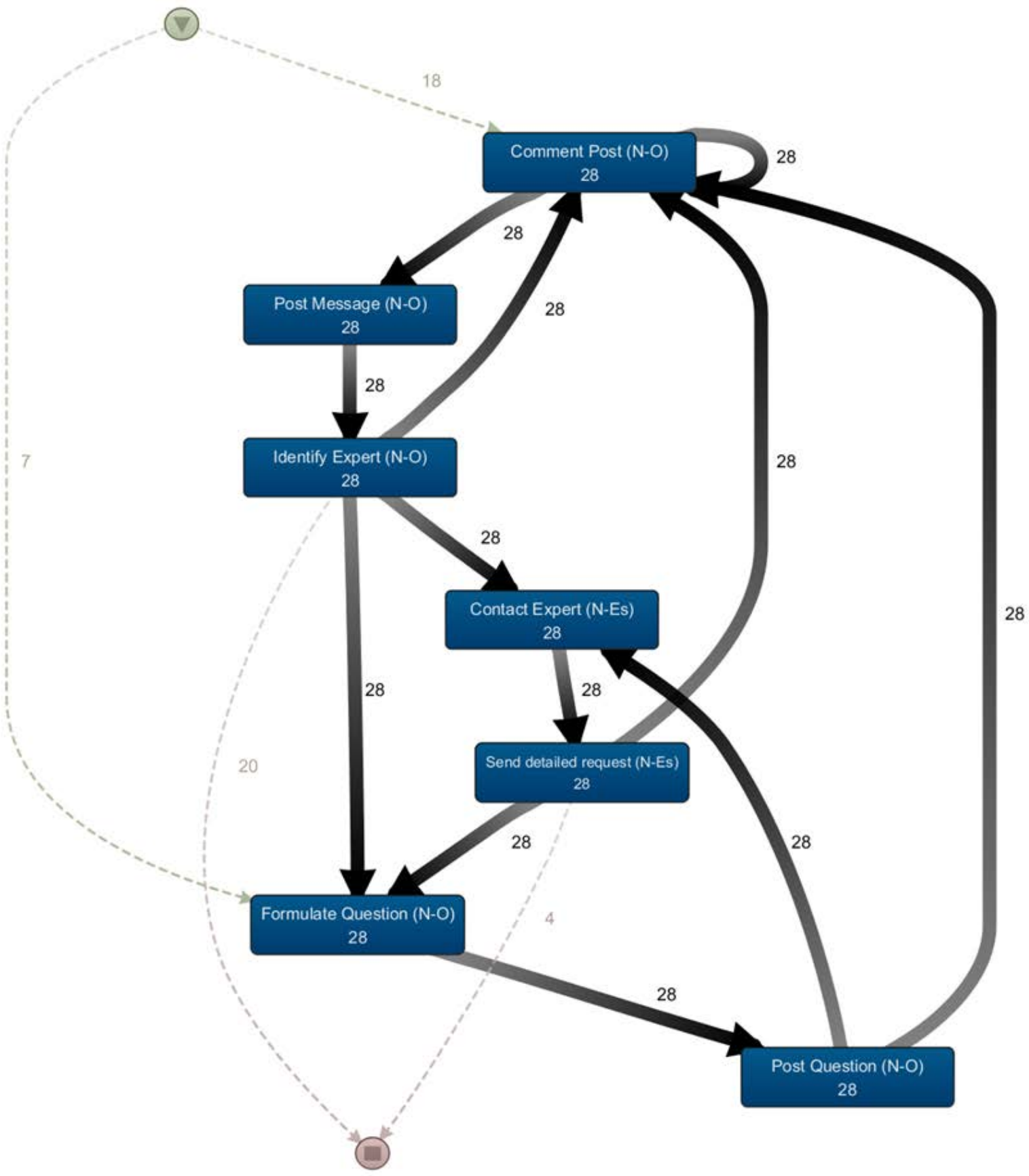

**Figure 8. Process map for Novice–Per frequency (Initiation Phase) in Internet Relay Chats**

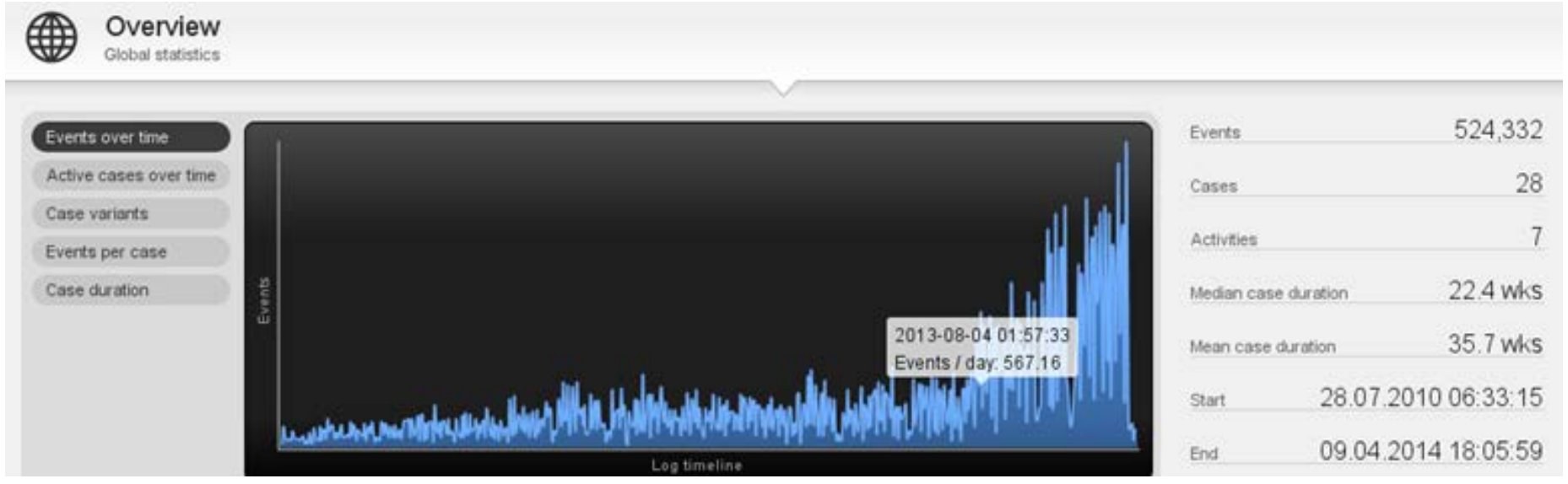

**Figure 9. Events over time for Novice**

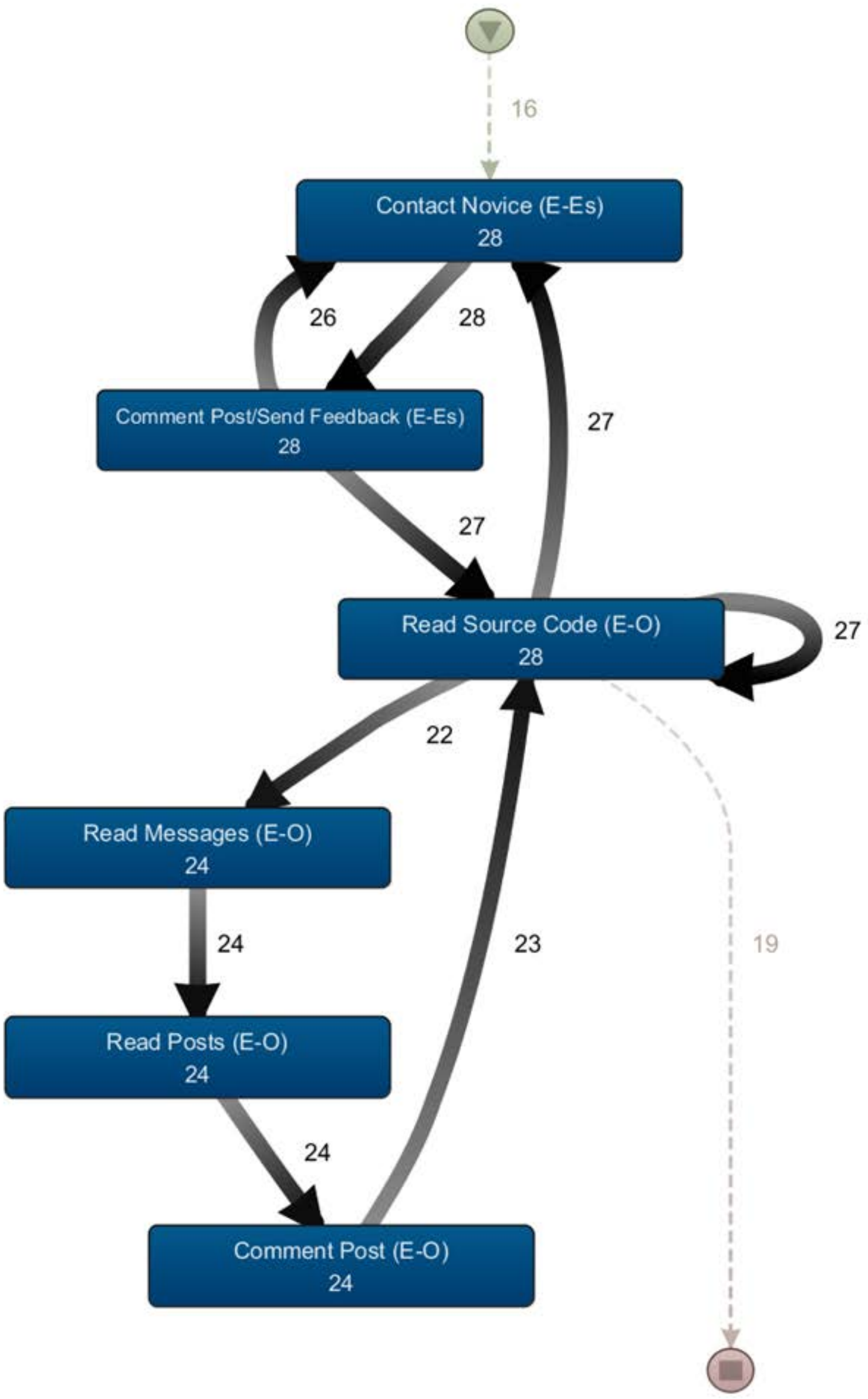

**Figure 10. Process map for Expert–Per frequency (Initiation Phase) in IRC messages**

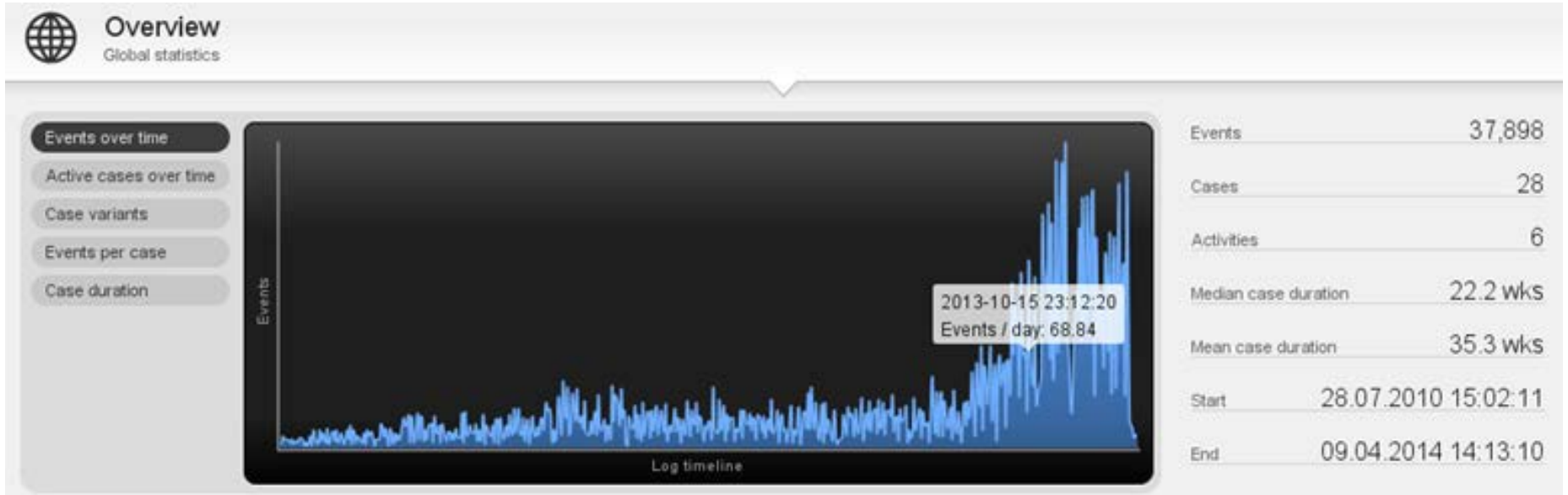

**Figure 11. Events over time for Expert**

Glott *et al.* [22] extensively describe the background of people involved in knowledge exchange, the types of software engineering skills that can be acquired in FLOSS communities but come short of explaining the possible sequence of pattern of occurrence. Moreover, Glott *et al.* [21] as well as Dillon *et al.* [14] discuss the potentials of using FLOSS principles to improve formal ICT education with limited or no direct empirical evidence from FLOSS repositories. Many others, including Fernandes *et al.* [13] critically provide insights that FLOSS members are aware that learning is taking place but are little aware of how this occurs.

## 5.2 Process Maps for the Progression Phase

In modeling this second phase of the learning process, we make use of the same dataset, given that the activities belonging to this phase can also be traced in IRC messages. We represent major statistical details that are most representative of the presence, impact and occurrence of learning activities in FLOSS over the chosen period of time through the process maps.

The Novice’s learning process map in Figure 12 indicates that, on IRC messages, the Novice starts in 14 cases with reporting bugs as a result of the Expert’s guidelines. Then, the Novice analyzes source code, comments on that code, and then either goes back to reporting bugs, or replies to posted questions, or provides/sends feedback, or posts questions. If posting questions, the Novice also posts further questions as part of reviewing posts or committed code, replies to posted questions and then goes back to analyze any relevant Source Code.

Regarding the Expert, we note in Figure 13 that the activities are organised in a number of intersecting cycles. The main path comprises two cycles and shows that, in some cases, the Expert starts sending a feedback to a request from the Novice, then reports a bug accordingly, replies to a post, reviews thread code, comments on code, analyzes Source Code, runs the code and then reviews the code, possibly reports further bugs and posts messages, and, finally, restarts the reviewing code cycle once again. Along the way, the Expert moves back and forth from cycle to cycle as needed.

Finally, it is important to note the increase in the involvement of the Expert in this phase comparing to the Initiation phase. Performing 12 activities out of a total of 21 activities, the Expert exemplifies the commitment and willingness to assist and help as demonstrated and reported in previous narrative findings [24-28].

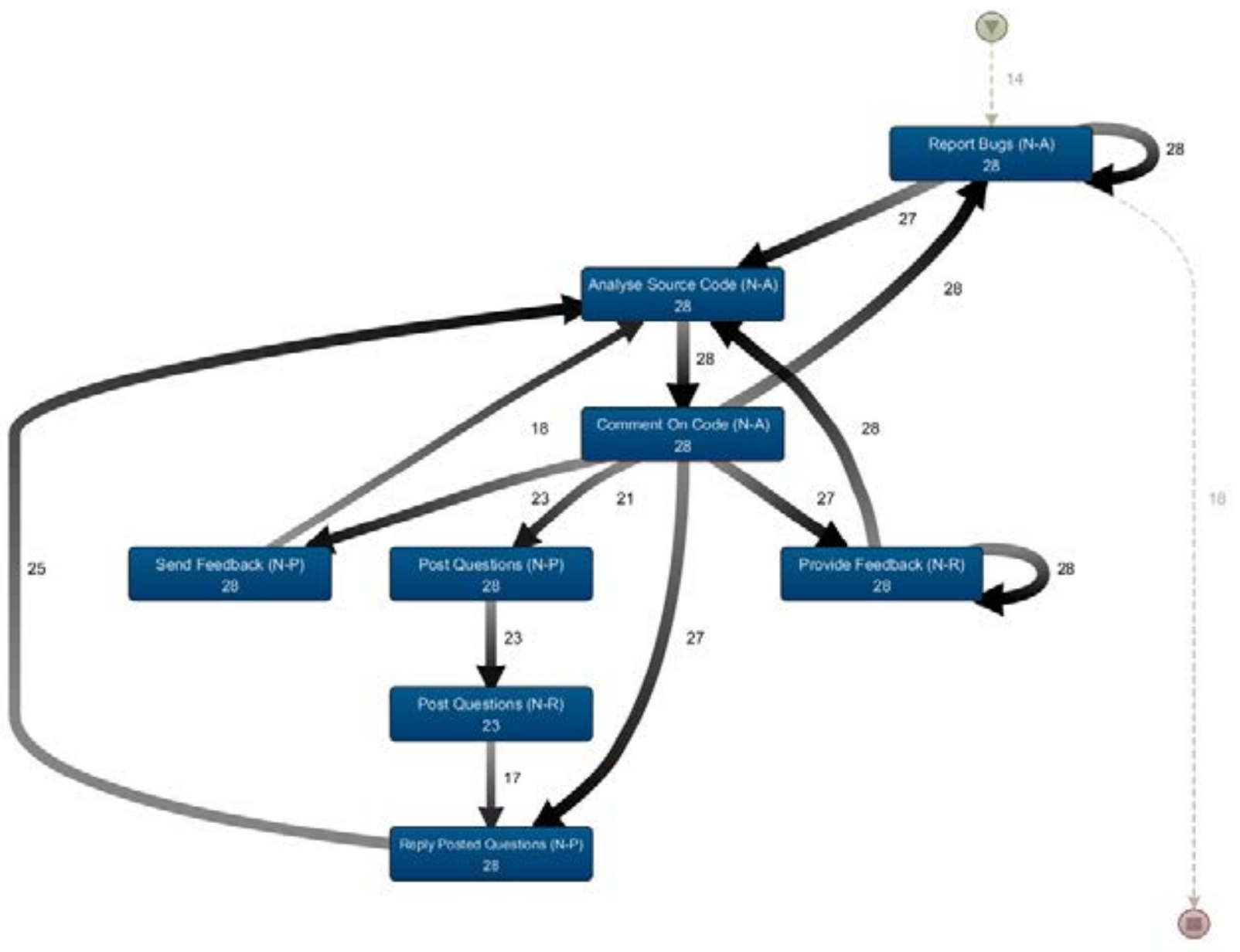

**Figure 12. . Process Map for Novice–Per frequency (Progression Phase) in IRC messages**

Figure 13. Process Map for Expert–Per frequency (Progression Phase) in IRC messages

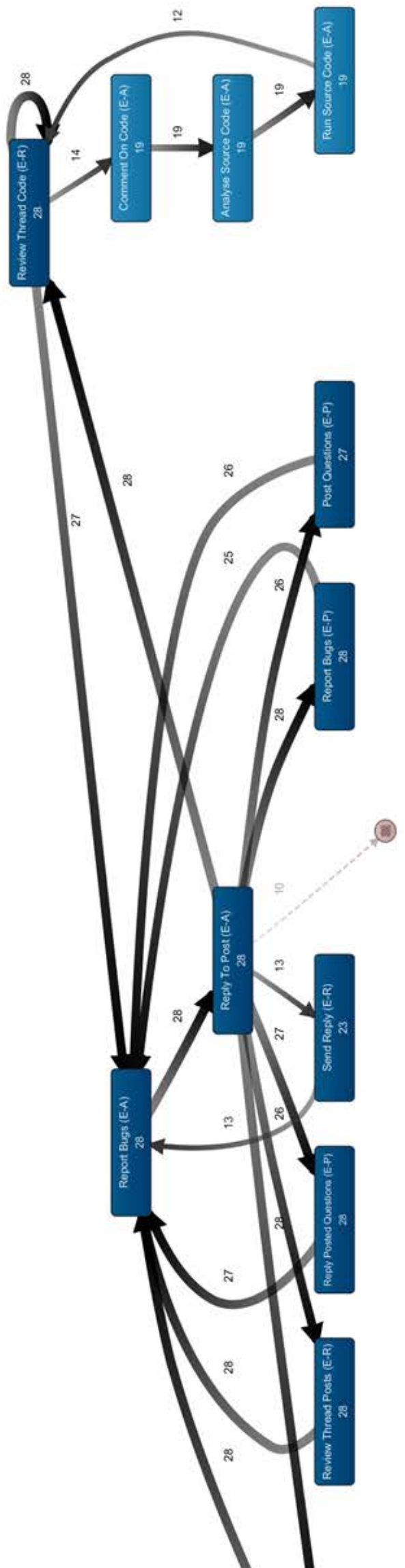

### 5.3 Process Maps for the Maturation Phase

The Maturation phase represents the degree at which the learning process has grown. This can evidently be seen through more advanced actions spanning from developing and committing pieces of code to reporting and fixing bugs. Hence, in this phase the number of activities sharply increases for both participants.

This phase continues to exemplify the extent to which people are eager to learn with the Novice executing most activities with a total of 506814 activities amounting to 46.19% of all executed activities at this point, in contrast with the Expert who intervenes at a lower rate of 42.04% with 461278 activities, ahead of people doing something other than exchanging knowledge with 129189 activities. The increase in the number of activities also justifies the increase in the size of the corresponding process maps that represent the patterns as executed by the Novice and Expert respectively in Figures 14 and 15.

Figure 14 traces the paths a Novice would follow while executing learning activities during Maturation phase. We note that on Source Code, the Novice performs a number of activities either in parallel or sequentially depending on the level of involvement in the process. The main paths we observe show that the Novice starts reviewing Source Code, then reports bugs as a result of the review, before performing a range of other activities spanning from analyzing discussions and posting comments on code to submitting code. Then the Novice can modify Source Code, if needed, thus fixing bugs, and, finally, provide the resultant feedback.

The involvement of the Expert is shown in Figure 15: the Expert starts analyzing Source Code as part of monitoring the Novice's progress. Then, the Expert comments on the code and subsequently either sends a reply or further reviews the Source Code and reports bugs. In this phase, we argue that the Expert follows the Novice's activities for monitoring purposes and therefore, the execution of activities comprises several alternative sequences.

**Figure 14. Process map for Novice–Per frequency (Maturation Phase) on Source Code**

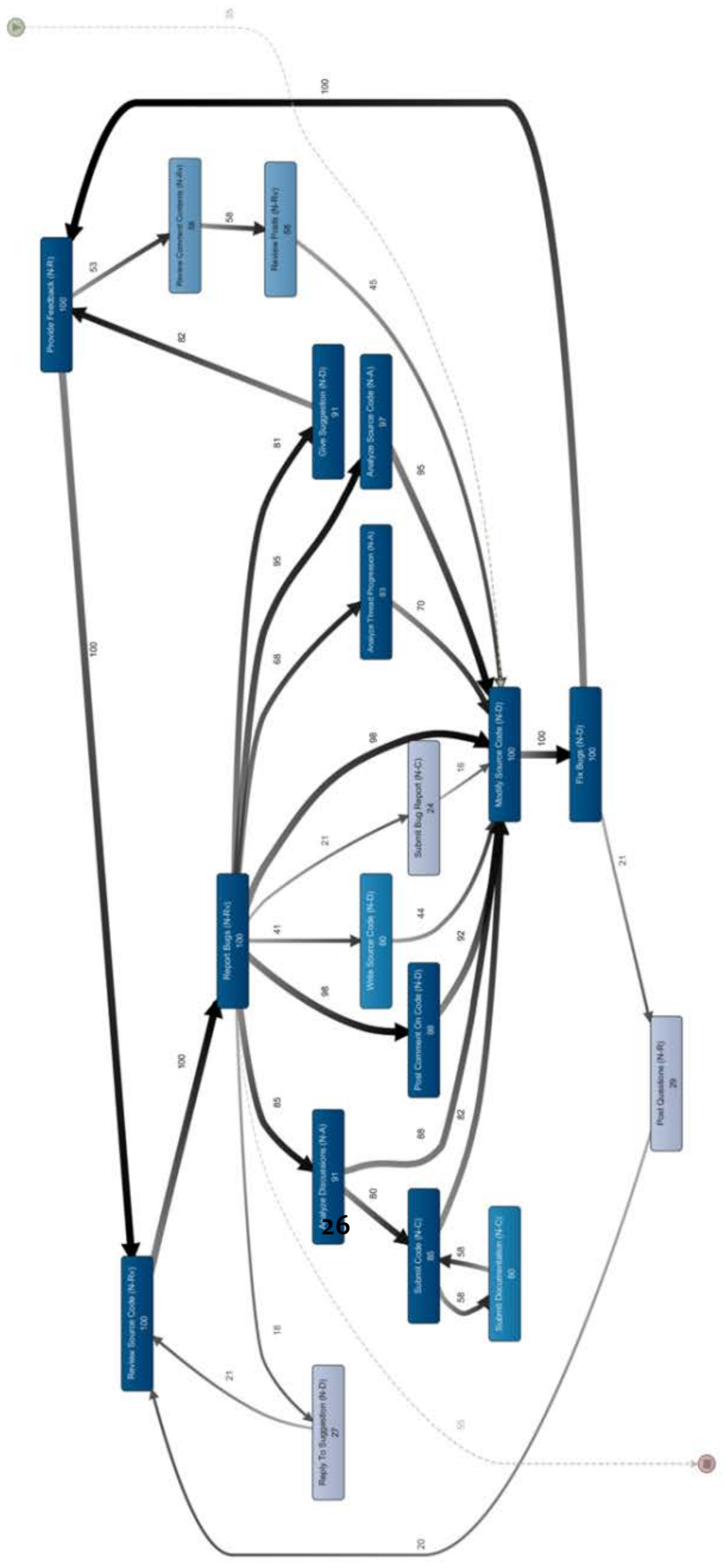

**Figure 15. Process map for Expert–Per frequency (Maturation Phase) on Source Code**

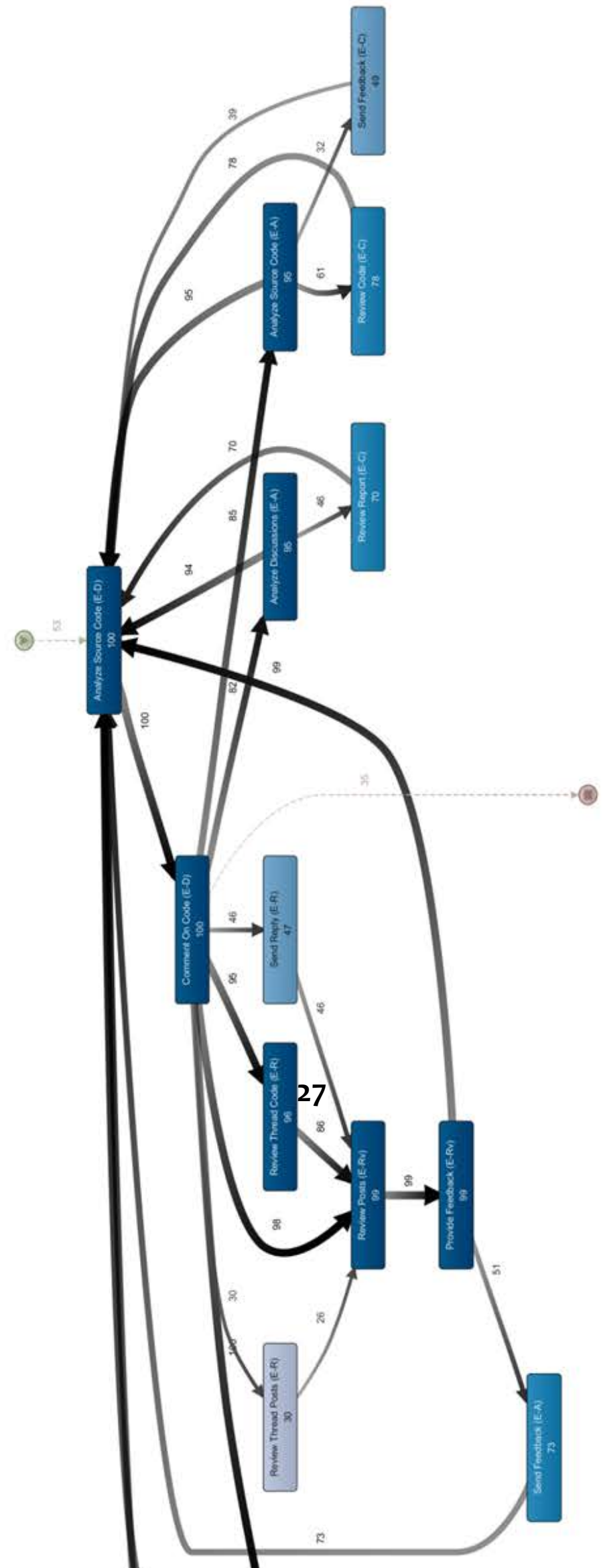

## 6 Conclusion

FLOSS communities can nowadays be considered as potential learning environments. Numerous studies conducted in this regard suggest that these environments present learning opportunities to users and members of these online platforms. While such studies provide invaluable insights, their results are mostly based on either surveys, observation reports [10, 14-16, 21-22], stochastic or probability studies [29]. Moreover, the learning potentials offered by FLOSS environments open new possibilities for teaching Software Engineering courses in tertiary formal institutions [3, 7, 10-14]. In this regard, we can highlight a number of pilot studies aimed at evaluating the effectiveness of such an approach in traditional settings of learning [2-5, 14-15].
Furthermore, while specific insights regarding learning in FLOSS are provided in these studies, little is shown in terms of the manner in which learning takes place. To our knowledge, existing literature on learning in FLOSS has not provided insights that would indicate how the communication occurs, the sequential occurrence of learning activities and, more importantly, the level of occurrence of learning activities in FLOSS communities. This level of occurrence could be expressed by looking at the number of learning activities proportionally to the total number of activities that are recorded in repositories.

Glott *et al.* [22] extensively describe the background of people involved in knowledge exchange and the types of software engineering skills that can be acquired in FLOSS communities, but come short of explaining the possible sequence of pattern of occurrence. Moreover, Glott *et al.* [21] as well as Dillon *et al.* [14] discuss the potentials of applying FLOSS principles to improving formal ICT education with limited or no direct empirical evidence from FLOSS repositories. Many others, including Fernandes *et al.* [13], critically provide insights that FLOSS members are aware that learning takes place but with limited indication on how this occurs.

In this paper, we have proposed an approach based on Process Mining. Using a combination of semantic search, ontology and lexical linguistics, we construct Event logs that capture the learning behavior of both the Novice and Expert in FLOSS. In order to show how communication occurs as well as the inherent execution of learning activities, we introduce an exploration of learning processes as process maps. Process maps provide valuable information about a process to help management find ways to make the process better. The detailed step-by-step descriptions included in process maps provide a clear and concise blueprint for content.

The accomplishments of our results are twofold. Firstly, we provide a simplified description of learning processes through process mapping. The process maps contribute to easily depict a pictorial representation of the occurrence of learning patterns in FLOSS environments with no need for detailed explanations. Secondly, the quantitative results provide critical insights with regards to the level of occurrence of learning activities in FLOSS repositories. We note that learning activities do occur at a significant rate during messages exchange for both participants across the three phases. On Source Code as well, as the learning process matures, the involvement of both the Novice and Expert significantly increases across the Progression and Maturation phases.

Although existing literature is critical in guiding the study of learning behavior in FLOSS, in particular, with regard to the phases of the learning processes [18-19, 29], our results provide more tangible evidence on how these phases can be traced from actual recorded data. The details that our experiments have uncovered provide richer insights regarding both roles in the learning process. The process maps executed by both the Novice and Expert provide a richer description of learning activities than provided in the current literature. This constitutes a critical step towards explicating and detailing the steps FLOSS members go through as part of learning and, more importantly, the extent to which learning in FLOSS communities is a reality.

Finally, we would like to point out that there are many more FLOSS repositories that could be mined to undertake similar experiments; however, we believe that the datasets from Internet Relay Chat messages and Source Code include enough details that provide global insights on recorded activities in FLOSS environments. Specifically, IRC messages appeared to be the right choices for the detection of process maps as they represent the learning process through its first two phases. In such phases, the role of the Novice is the most active, justifying the high intensity for the quest for knowledge. Analogously, the Expert becomes progressively more and more involved from Initiation to Progression phases. Regarding the last phase of the learning process, the choice of the Source Code dataset largely explains how learning occurs at this stage. Evidence suggests the commitment of the Novice to seek answers and interact as much as possible in strengthening the acquired skills.